\documentclass[aps,twocolumn,showpacs]{revtex4}

\usepackage{bm}
\usepackage{graphicx}
\usepackage{amsmath}
\usepackage{amssymb}

\def\lan{\langle}
\def\ran{\rangle}
\def\r{{\bm r}}

\begin{document}

\title{Dynamical behavior of a lattice glass model on a random graph: comparison with Mode Coupling Theory}

\author{A. de Candia$^{a,b,c}$, M. Mauro$^d$, A. Coniglio$^{a,b,c}$}

\affiliation{%
$^a$Dipartimento di Scienze Fisiche, Universit\`a di Napoli Federico II\\
$^b$INFN, Sezione di Napoli\\
$^c$CNR-INFM Coherentia, Napoli\\
$^d$PROMETE S.r.l., Business Innovation Center di Citt\`a della Scienza, Napoli
}


\begin{abstract}
We study the dynamical behavior of a lattice model of glass former on a random graph,
where no corrections to the mean field description are expected.
We find that the behavior of dynamical correlation functions and
dynamical susceptibility are consistent with the quantitative predictions of the Mode Coupling Theory of the glass transition.
\end{abstract}

\maketitle

The glass transition is characterized by a dramatic rise of the relaxation times of the system,
over a small interval of temperatures or densities. 
It has long been supposed that this phenomenon is connected to the growing of a cooperativity length in the dynamics \cite{adamgibbs,kirkwol}.
This idea suggests the existence of dynamical heterogeneities in supercooled liquids, that is groups of particles
whose motion is spatially correlated, on a scale that becomes large as the glass transition is approached.
Such heterogeneities were identified in experiments \cite{cicerone} and simulations \cite{donati}.
A way to quantify this dynamical heterogeneities is in terms of the four point (dynamical) susceptibility $\chi_4(t)$ \cite{frapa,fdpg},
defined as $\chi_4(t)=V\left[\lan \phi(t)^2\ran-\lan\phi(t)\ran^2\right]$, where $\phi(t)=\rho(\r,0)\rho(\r,t)-\lan\rho\ran^2$,
and $\rho(\r,t)$ is the density at position $\r$ and time $t$.
The susceptibility $\chi_4(t)$ is the volume integral of a four point correlator,
$\chi_4(t)=\int G_4(\r,t) d^3\r$, where
\begin{align}
G_4(\r,t)=&\lan\rho(0,0)\rho(0,t)\rho(\r,0)\rho(\r,t)\ran
\nonumber\\
&-\lan\rho(0,0)\rho(0,t)\ran\lan\rho(\r,0)\rho(\r,t)\ran,
\end{align}
measures the correlations in space of local time correlation functions. The function $\chi_4(t)$ has been computed in
many model glass formers, and indeed it displays a maximum at a time $t^\ast\sim\tau_\alpha$, where
$\tau_\alpha$ is the relaxation time of density fluctuations, with the maximum $\chi_4(t^\ast)$ growing
when approaching the transition, signaling the growing of the range of $G_4(\r,t^\ast)$,
that is of dynamical correlations.

A mean field level description of the glass transition is the Mode Coupling Theory (MCT) \cite{mct}, which makes many predictions
on the form of the relaxation near the transition. In particular it predicts a power law divergence, $\tau_\alpha\sim |T-T_c|^{-\gamma}$,
of the relaxation times.
It was recently shown \cite{birbou} that, within the mean field description of MCT,
the time $t^\ast$ where $\chi_4(t)$ has a maximum diverges with the same exponent of the structural relaxation time,
$t^\ast\sim\tau_\alpha\sim |T-T_c|^{-\gamma}$. Furthermore, the value of the maximum diverges as
$\chi_4(t^\ast)\sim {\tau^\ast}^x$, with $x=1/\gamma$.
One also finds that, in the early and late $\beta$ regime, 
the dynamical susceptibility behave as $\chi_4(t)\sim t^\mu$, respectively with $\mu=a$ and $\mu=b$ \cite{twbbb},
where $a$ and $b$ are the standard MCT exponents that describe the approach and departure from the {\em plateau}.

To check these results using molecular dynamics on a realistic model of glass forming liquid, like a Lennard-Jones mixture,
is not an easy task for two reasons.
The first is that simulations are computationally heavy, and it is difficult to reach very long simulation times. The second is that
the predicted Mode Coupling transition is smeared out by corrections to mean field, the so called hopping processes, so that
near the transition a crossover to a different regime is observed.
A class of models in which the MCT is exact are some spin glass models with a discontinous transition, for example the $p$-spin model \cite{pspin}.
However, such models are characterized by quenched disorder, that gives rise to strong finite size effects, due to sample to sample fluctuations
of the critical temperature, as pointed out in a recent paper \cite{birnew}.
For this reason, in such models the asymptotic critical behaviour, that is expected only in a very small region around $T_c$,
is not easily observed.

In this paper, we study the dynamical behavior of a lattice model of glass forming liquids
on the random graph (or Bethe lattice).
Being defined on lattice, the model is much less computationally expensive than more realistic models.
Furthermore,
due to the tree-like nature of the graph, we expect the critical behaviour to be described correctly by Mode Coupling Theory.
Finally, the model has no quenched disorder (apart from the randomness of the graph), so that it is reasonable to expect not
too large sample to sample fluctuations.
The model is defined as follows \cite{ulm}. The space is partitioned in regular cells, such that
not more than one particle can have its center of mass inside the cell.
The position of the particle inside the cell is discretized, so that it can assume a finite number $q$ of positions.
The model is therefore described by the following Hamiltonian:
\begin{equation}
{\cal H}=\sum_{\langle ij\rangle} n_i n_j \phi_{ij}(\sigma_i,\sigma_j)-\mu\sum_in_i,
\end{equation}
where $n_i=0,1$ represents whether the $i$-th cell is occupied or not, $\sigma_i=1\ldots q$ represents the position of the particle
inside the cell, and $\phi_{ij}(\sigma_i,\sigma_j)$ the interaction between two particles in neighboring cells $i$ and $j$, with
positions $\sigma_i$ and $\sigma_j$.

\begin{figure}[ht]
\begin{center}
a)
\includegraphics[width=2.5cm]{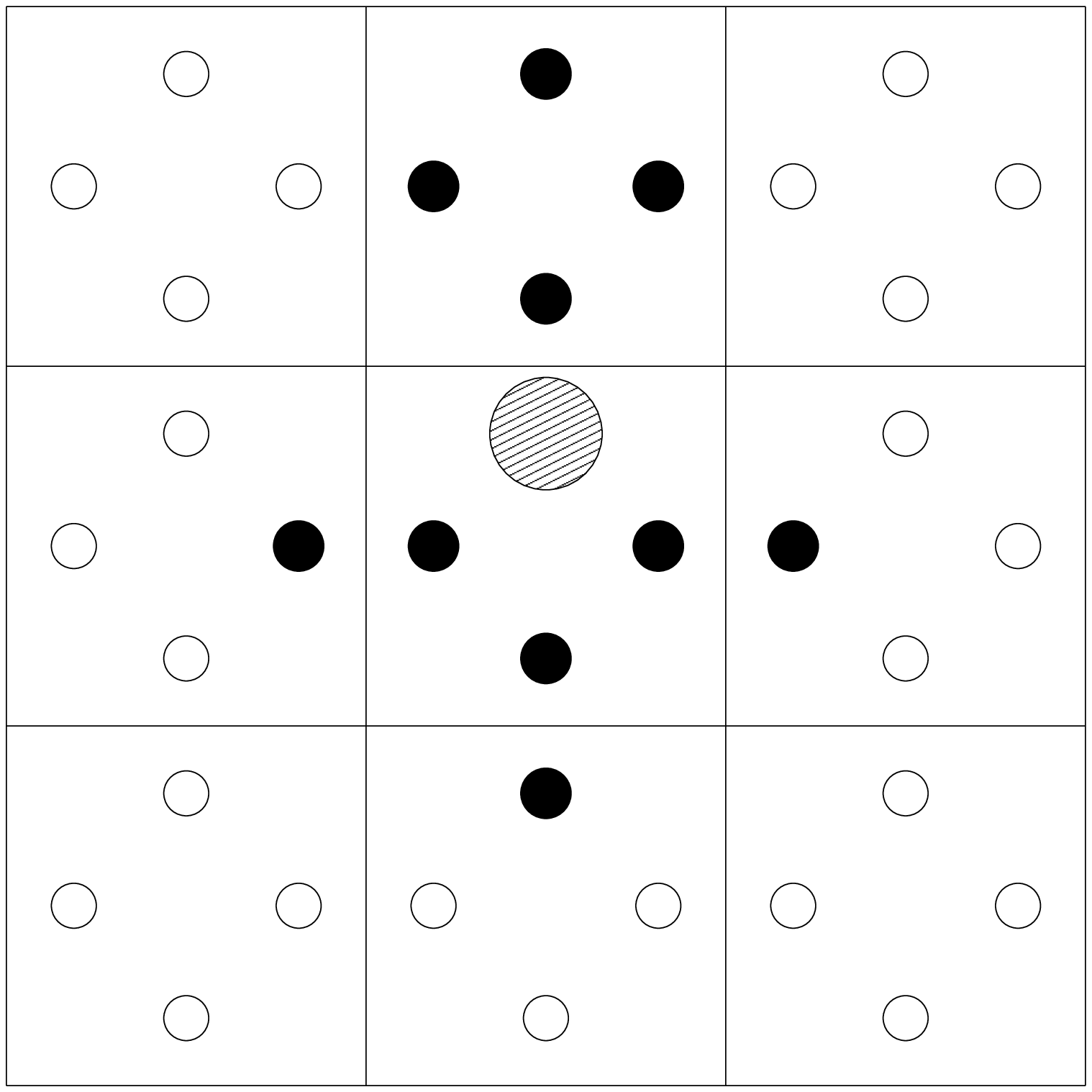}
\qquad
b)
\includegraphics[width=3cm]{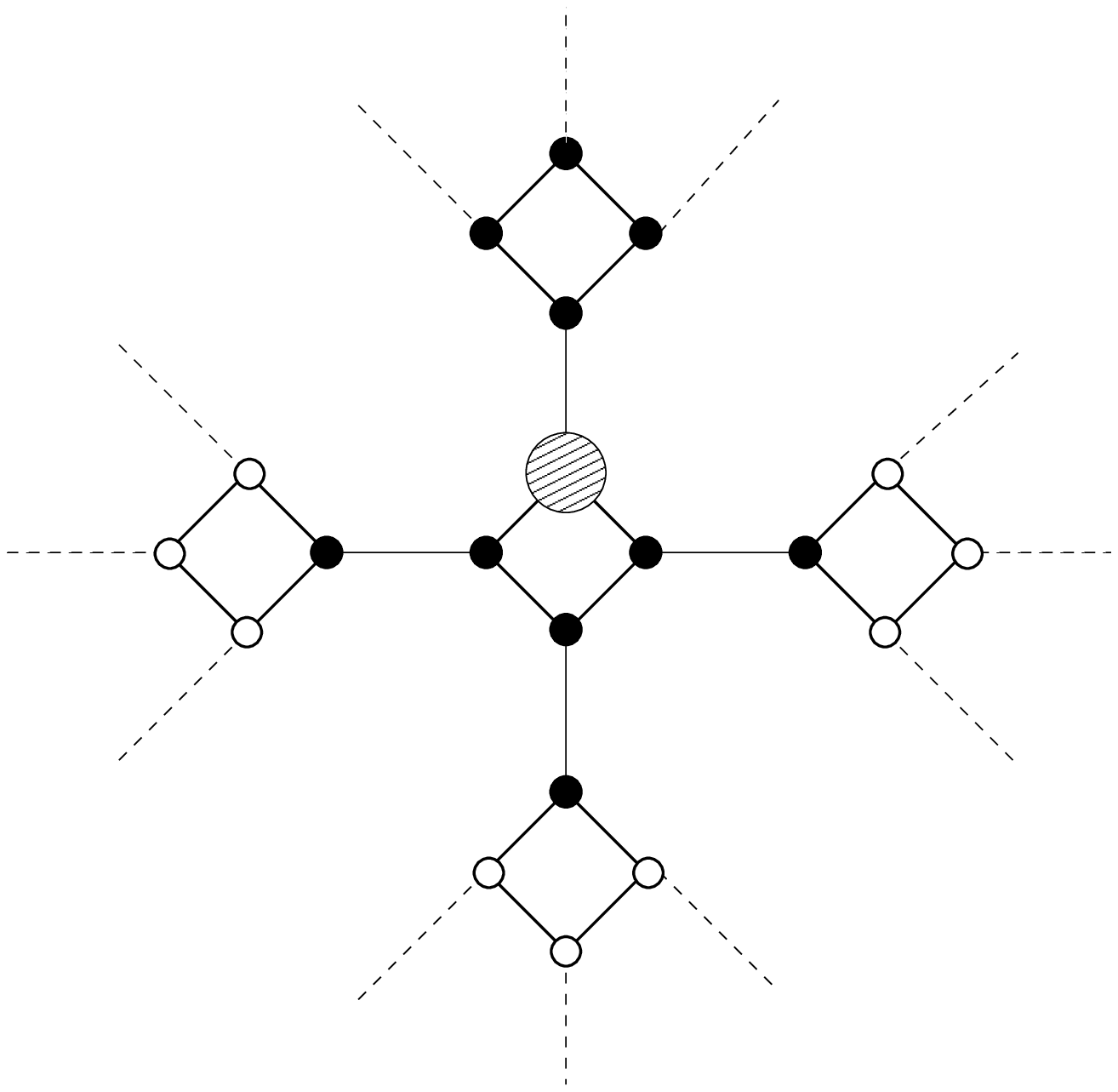}
\end{center}
\caption{%
a) The model in two dimensions: the space is partitioned in square cells, and each cell can be occupied by at most one particle
in one of four positions (little circles). A particle in a given position (big shaded circle) forbids the presence of another particle in the
positions colored in black.
b) The model on the random graph with $z=4$.
}
\label{figlatt}
\end{figure}

A particularly simple realization of the model is shown in Fig.\ \ref{figlatt}a for two dimensions. In this case
the space is partitioned in square cells, with $q=4$ positions per cell,
and the interaction $\phi_{ij}(\sigma_i,\sigma_j)$ is zero, if the positions $\sigma_i$ and $\sigma_j$ are ``compatible'',
infinite otherwise.
In Fig.\ \ref{figlatt}b the same model is defined on the random graph, with the same value $z=4$.
The graph in this case is extracted randomly from the set of all regular graphs where each site has exactly $z$ neighbors \cite{nota1}.
This kind of graph is  locally tree-like, as there are no cycles up to a distance scaling as $\log N$ from a typical site.
However, the presence of long cycles induces frustration and ensures that the system is statistically homogeneous.
The mean field approach is exact on such graphs, due to the fact that the local fields acting on a site are uncorrelated
in the thermodynamic limit.

\begin{figure}[ht]
\begin{center}
\includegraphics[width=3.5cm]{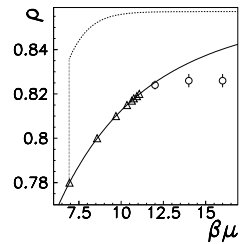}
\end{center}
\caption{%
Density $\rho$ as a function of $\beta\mu$, calculated analytically (lines and circles)
on the random graph with $z=6$ (from Ref.\ \cite{ulm}).
Solid line represents the liquid phase, while the dashed line the crystalline phase.
Circles: density in the 1-step RSB {\em ansatz}.
Triangles: chemical potential as a function of the density, evaluated in the simulations performed in the present paper on a $N=10^5$ random graph.
}
\label{figrho}
\end{figure}

In Refs.\ \cite{ulm} the thermodynamics of the model on the random graph was studied analytically.
Using a 1-step replica symmetry breaking (RSB) {\em ansatz}, it was found \cite{ulm} that in the case $z=6$ the model
has a ``dynamical transition'' at a density $\rho_d=0.808$, where self-consistent 1-RSB solutions appear,
signaling the appearing of an extensive number of metastable states.
The transition is called dynamical, because it is usually connected to a MCT-like dynamical freezing, as it happens
in the $p$-spin model \cite{pspin}. At a higher density $\rho_K=0.822$, the free energy of the 1-RSB solution equals the free energy
of the replica symmetric solution, signaling the vanishing of configurational entropy, and the presence of a
thermodynamical Kautzmann transition.
In Fig.\ \ref{figrho}, the solid and dashed lines represent the density evaluated analytically in the replica symmetric approximation,
respectively in the liquid and solid (crystalline) phase, while the circles represent the density in the 1-step RSB {\em ansatz}.

In the present paper, we investigate the dynamical behavior of the model with $z=6$, corresponding to the three dimensional case.
We perform two types of Monte Carlo dynamics on the system, canonical (fixed density) and grand-canonical (variable density).
The grand-canonical dynamics is given by the following algorithm.

(1) Pick up a site at random. If the site is occupied by a particle,
then pick up at random one of the $z$ nearest positions (the other $z-1$ positions inside the same site
and the nearest position of the nearest site) and, if it is allowed by the hard core constraints,
move the particle in the new position.

(2) Pick up a site and a position at random. If the position is occupied by a
particle, destroy the particle with probability $\exp(-\mu/T)$.
If the position is empty, create a new particle if it is allowed by the hard core constraints.

(3) Advance the time by $1/N$, where $N$ is the number of sites.

In the canonical dynamics step 2 is missing.
Near the dynamical transition, where the probability to make a move is small,
it is convenient to implement smart Monte Carlo algorithms like the ``$N$-fold way'' \cite{nfold},
that save computer time leaving the dynamics unchanged.

We consider graphs composed by $N=10^5$ sites (cells), each having $z=6$ neighboring sites.
First of all, we extract a regular random graph with fixed connectivity $z$. Starting from the empty graph, we perform
grand-canonical dynamics with a slow cooling rate $d\mu/dt$, until we reach the desired density. Therefore, we perform
canonical dynamics, skipping at least $10^9$ Monte Carlo steps for thermalization.
All the quantities are averaged over 16 different realizations of the random graph.
The relation between density and chemical potential can be evaluated in the canonical dynamics by $\beta\mu=\ln(N_p/\lan N_f\ran)$,
where $N_p$ is the number of particles, and $\lan N_f\ran$ is the mean number of free positions, that is of positions where a particle might be
created without violating geometric constraints. In Fig.\ \ref{figrho}, the correspondence between density and chemical potential in
simulations is shown by the triangles. The values fall nicely on the curve corresponding to replica symmetry, indicating that
the system is in the equilibrium liquid phase for all density considered.

\begin{figure}[ht]
\begin{center}
a)\includegraphics[width=4.2cm]{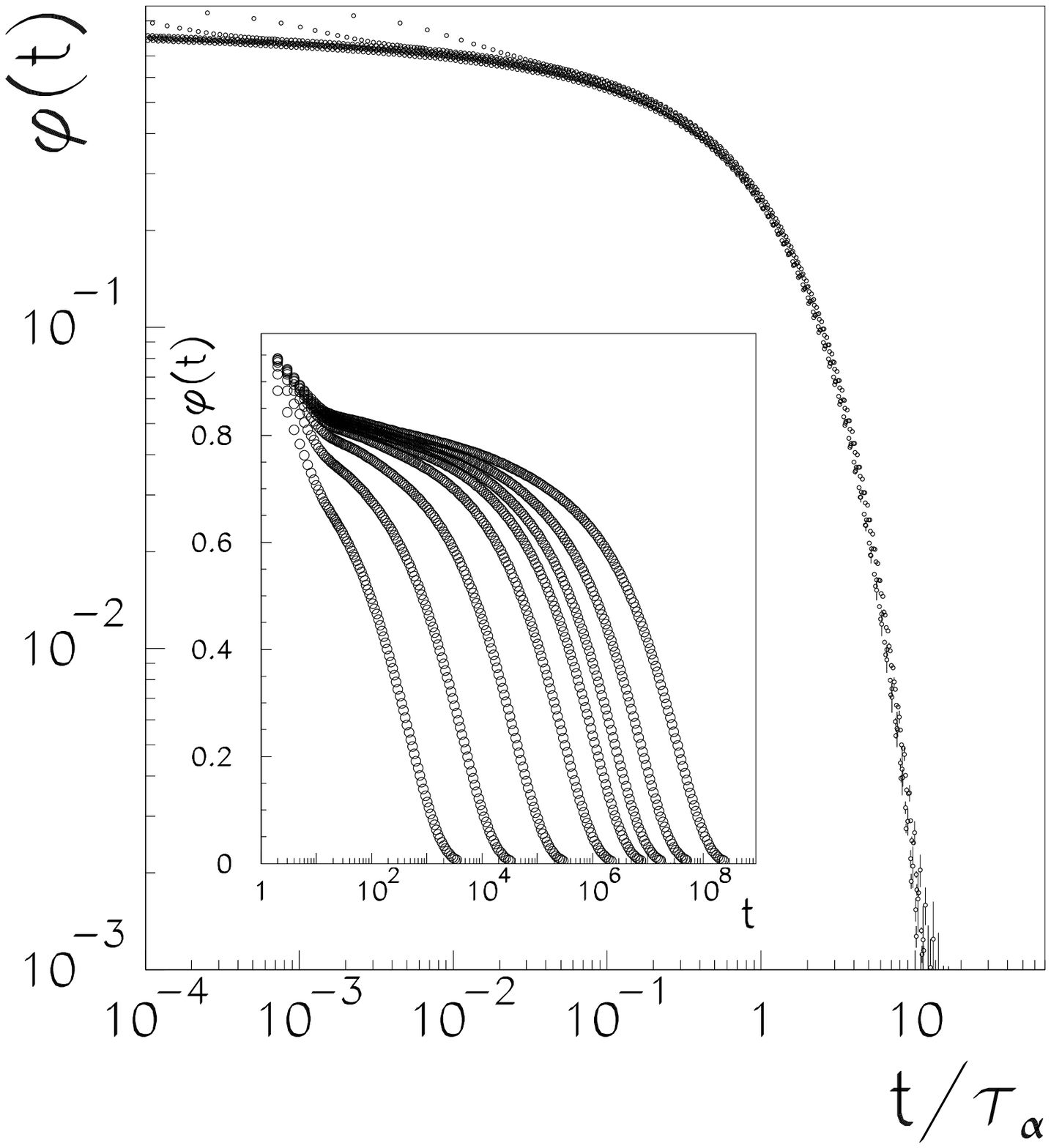}\\
b)\includegraphics[width=4.2cm]{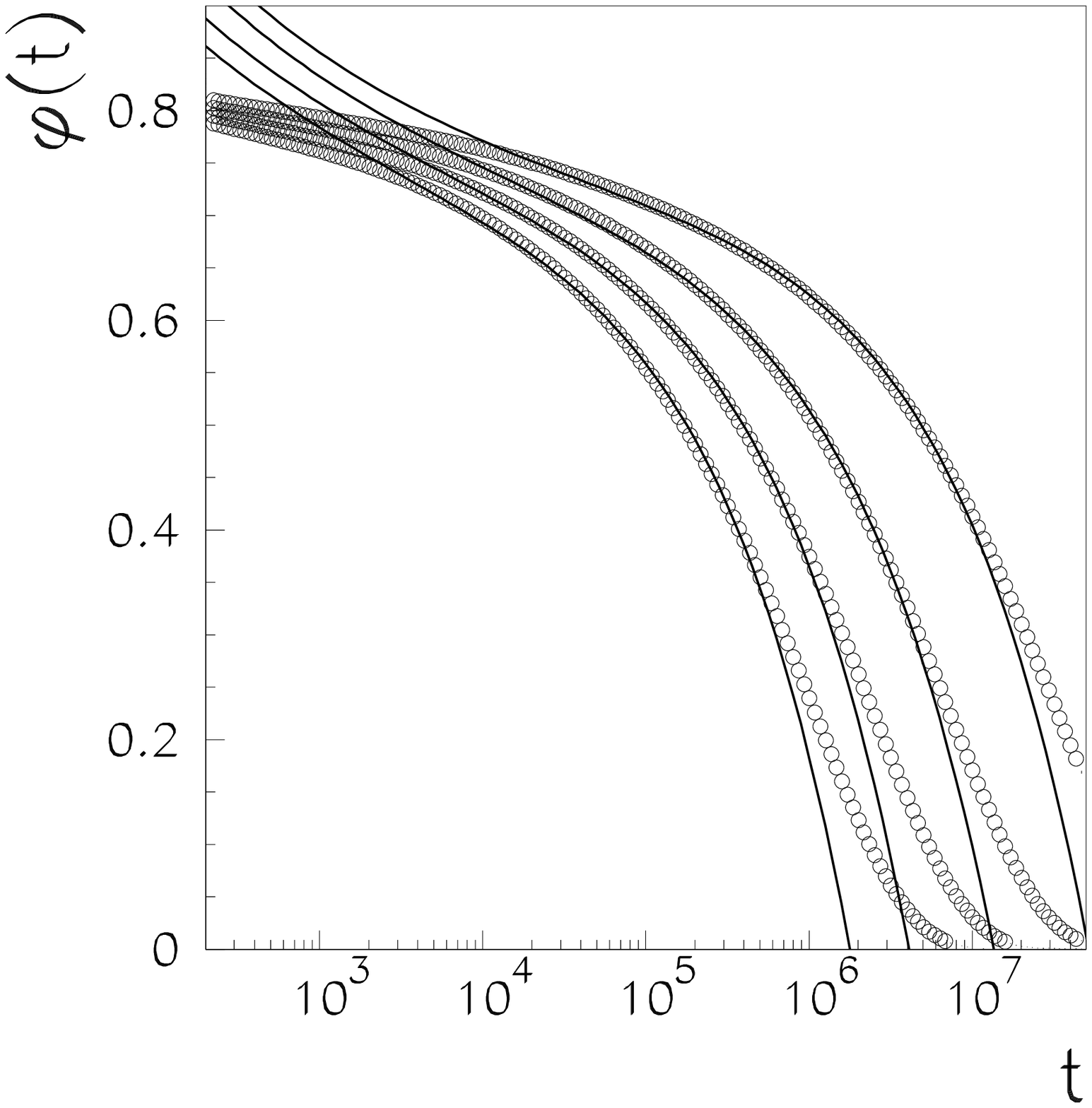}
\end{center}
\caption{a) Density-density relaxation function $\phi(t)$ as a function of $t/\tau_\alpha$, for densities (from left to right) $\rho=0.78$,
0.80, 0.81, 0.815, 0.817, 0.818, 0.819, 0.82. Inset: $\phi(t)$ as a function of $t$.
b) Fit of the function $\phi(t)$ for the four densities nearest to the transition with the function defined in Eq.\ (\ref{eq_gotze}).
}
\label{figrelax}
\end{figure}

\begin{figure}[ht]
\begin{center}
\includegraphics[width=4.2cm]{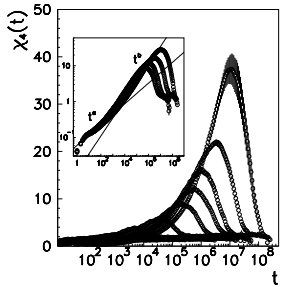}
\end{center}
\caption{Dynamical susceptibility $\chi_4(t)$  as a function of time, for the same densities as Fig.\ \ref{figrelax}a.
Inset: The same plot in logarithmic scale, for the four densities nearest to the transition.
The straight lines are the power laws $t^\mu$ with exponents $\mu=0.29$ and $\mu=0.50$,
corresponding to exponents $a$ and $b$ of MCT.
}
\label{figchi}
\end{figure}

\begin{figure}[ht]
\begin{center}
a)\includegraphics[width=4.2cm]{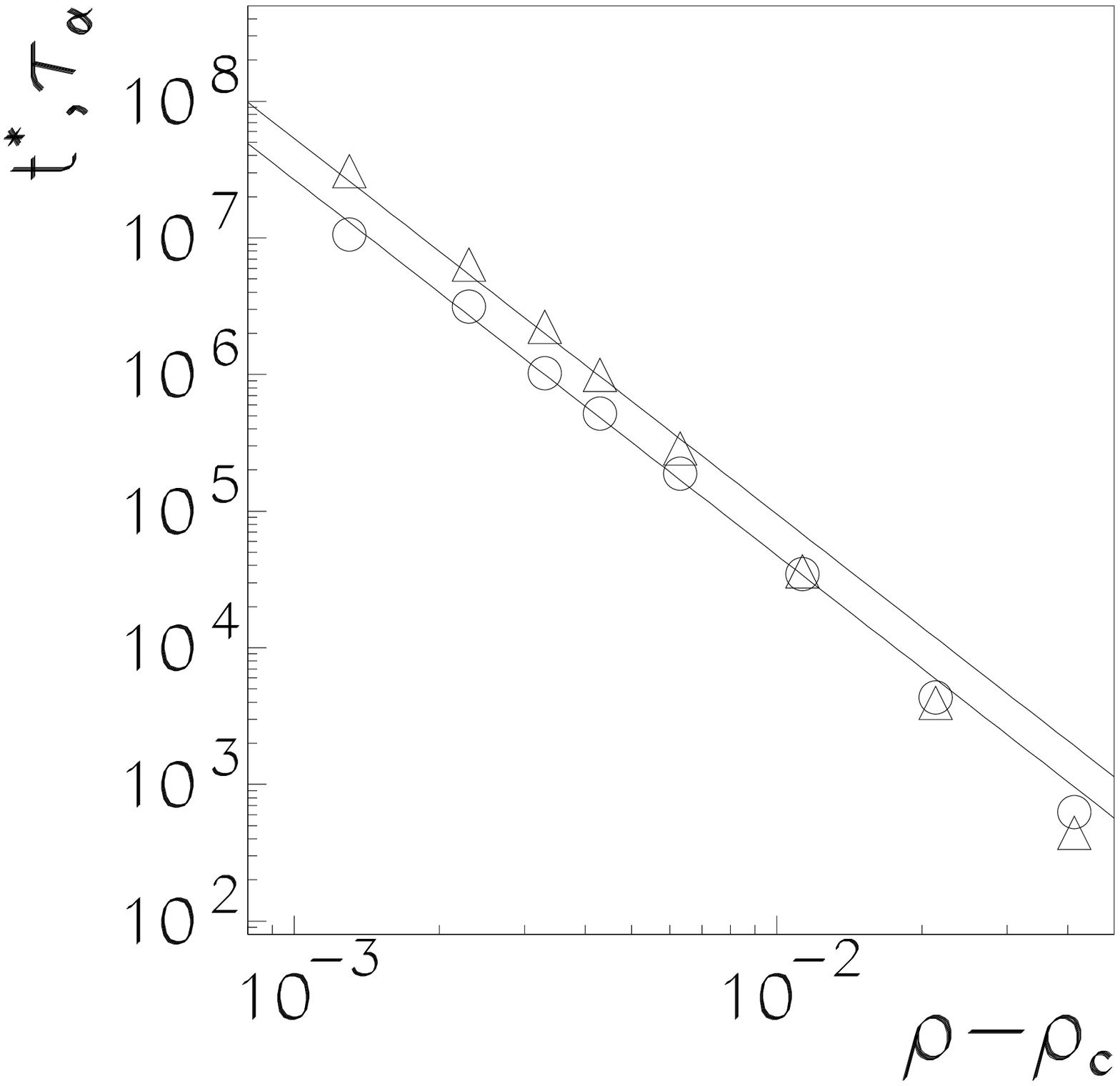}\\
b)\includegraphics[width=4.2cm]{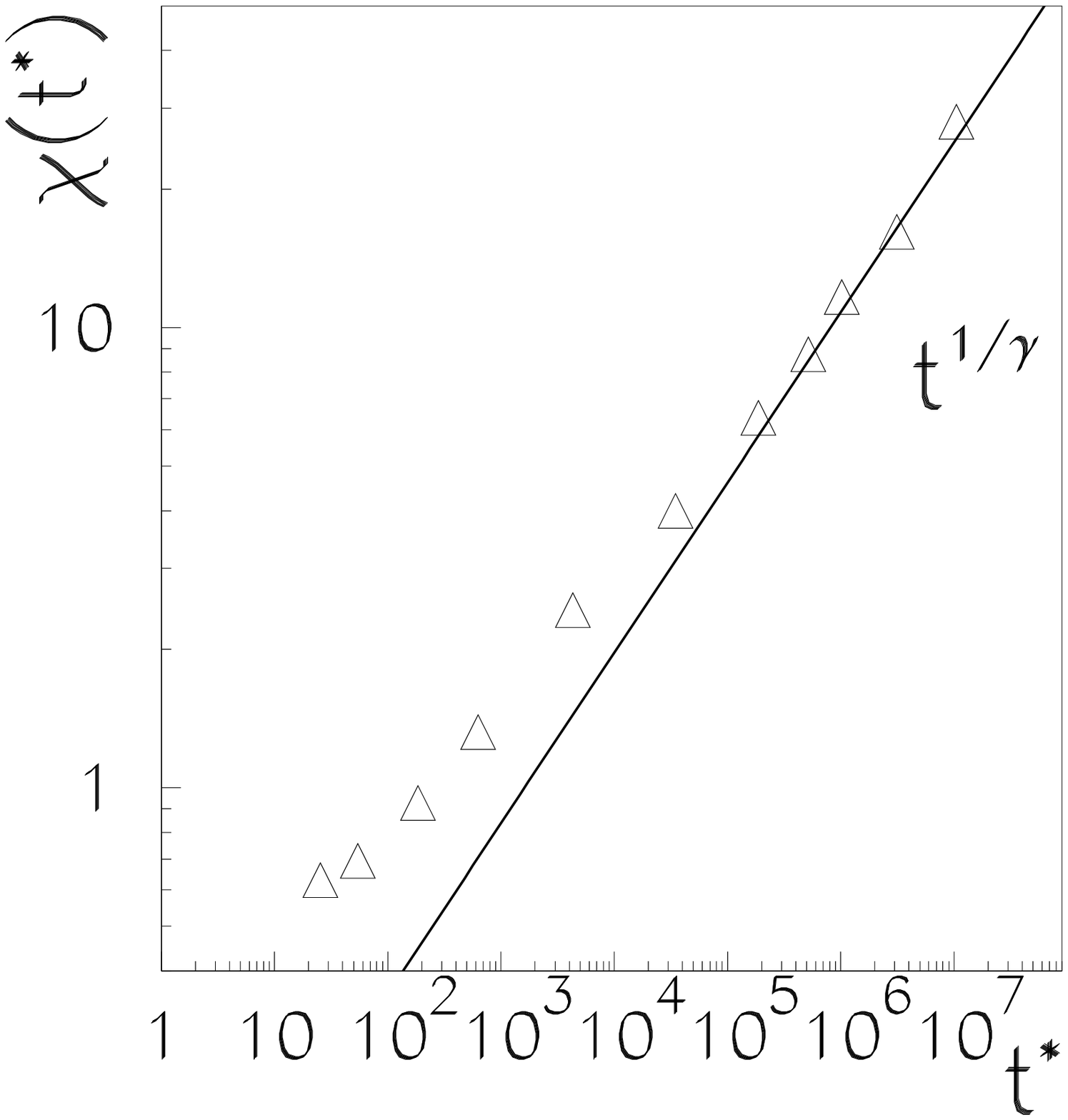}
\end{center}
\caption{a) Triangles: relaxation times $\tau_\alpha$, defined as the time given by the fit $e^{-(t/\tau_\alpha)^\beta}$.
Circles: time $t^\ast$ of the maximum of the dynamical susceptibility.
Lines: fitting functions $a|\rho-\rho_c|^{-\gamma}$, with $\rho_c=0.821\pm 0.001$ and $\gamma=2.75\pm 0.05$.
b) Maximum $\chi(t^\ast)$ of the dynamical susceptibility as a function of the time $t^\ast$.
The straight line is a power law $\chi(t^\ast)\sim {t^\ast}^x$ with exponent $x=0.36$, equal within errors to $1/\gamma$.
}
\label{figtau}
\end{figure}

We then studied the dynamical functions. For each value of the density, we computed the quantity
\begin{equation}
\phi(t)=\frac{\frac{z}{N}\sum_{i=1}^{N}n_i(0)n_i(t)\delta_{\sigma_i(0),\sigma_i(t)}-\rho^2}{\rho(z-\rho)}
\end{equation}
as a function of time, where $n_i(t)=0,1$ if site $i$ is empty or occupied at time $t$,
$\sigma_i(t)=0\ldots5$ is the position of the particle inside site $i$ at time $t$, and $\rho$ is the particle density.
The density-density relaxation function of the system is defined as $[\lan\phi(t)\ran]$,
where $\lan\cdots\ran$ is the (equilibrium) time average, while $[\cdots]$ is the average
over the graph configurations.
The result is plotted in the Inset of Fig.\ \ref{figrelax}a. We have fitted the long time tail of the functions with the
Kohlrausch-Williams-Watts stretched exponential function $fe^{-(t/\tau_\alpha)^\beta}$,
and found that $f$ and $\beta$ are nearly constant for different densities, ranging respectively
between 0.66 and 0.70, and between 0.74 and 0.77.
This shows that the time-temperature superposition principle of MCT is well verified in the whole range of densities considered,
as showed in the main frame of Fig.\ \ref{figrelax}a.

The dynamical susceptibility is defined as
\begin{equation}
\chi_4(t)=N\left[\lan \phi(t)^2\ran-\lan \phi(t)\ran^2\right].
\label{defchi}
\end{equation}
Fig.\ \ref{figchi} shows the behavior of $\chi_4(t)$. As expected, it has a maximum at a time $t^\ast$,
and both the time $t^\ast$ and the maximum $\chi_4(t^\ast)$ grow when approaching the transition.
Note that the long time limit of $\chi_4(t)$ is connected to a spin glass (static) susceptibility,
$\lim_{t\to\infty}\chi_4(t)=\chi_{\text{SG}}$, where
\begin{equation}
\chi_{\text{SG}}=\frac{z^3}{\rho^2(z-\rho)^2}\sum_{i,\sigma}\left(\langle n_0n_i\delta_{\sigma_0,0}\delta_{\sigma_i,\sigma}\rangle
-\frac{\rho^2}{z^2}\right)^2,
\end{equation}
and $\langle n_0n_i\delta_{\sigma_0,0}\delta_{\sigma_i,\sigma}\rangle$ is the probability
that there is a particle in site $0$ and position $\sigma_0=0$, and a particle in site $i$ and position $\sigma_i=\sigma$.
As it is apparent from Fig.\ \ref{figchi}, the spin glass susceptibility remains small when approaching the transition,
which means that static correlations between particles remain short ranged.

The relaxation times $\tau_\alpha$, extracted from the fit of $[\lan\phi(t)\ran]$ for all the densities considered,
and the time $t^\ast$ of the maximum of the dynamical susceptibility $\chi_4(t)$, are plotted in Fig.\ \ref{figtau}a
respectively as triangles and circles. Both $\tau_\alpha$ and $t^\ast$ can be fitted with a power law,
with the same (within errors) critical density $\rho_c=0.821\pm 0.001$ \cite{nota_rho}, and exponent $\gamma=2.75\pm 0.05$.
Note however that the critical behavior is observed only very near to the transition, for $|\rho-\rho_c|<0.005$,
as anticipated in Ref.\ \cite{birnew}.

From the value of $\gamma$, we can extract the values of the exponents $a$ and $b$ that describe the approach and departure from
the {\em plateau} and the exponent parameter $\lambda$ \cite{mct},
\begin{equation}
\gamma=\frac{1}{2a}+\frac{1}{2b},\qquad
\frac{\Gamma(1-a)^2}{\Gamma(1-2a)}=\frac{\Gamma(1+b)^2}{\Gamma(1+2b)}=\lambda.
\end{equation}
The corresponding values are $a=0.287\pm 0.004$, $b=0.50\pm 0.01$, $\lambda=0.785\pm 0.006$.

The MCT of the glass transition predicts that in the $\beta$ regime the relaxation functions can be fitted with the function
\begin{equation}
\phi(t)=f+h\sqrt{|\sigma|}g(t/t_\sigma),
\label{eq_gotze}
\end{equation}
where $t_\sigma=|\sigma|^{-\frac{1}{2a}}$, $\sigma=\rho-\rho_c$, and $g(x)$
is a specific function depending on the exponent parameter $\lambda$ (for more details see Ref.\ \cite{mct}).
We have fitted the relaxation functions with the function (\ref{eq_gotze}), for the four densities nearest to the transition,
fixing $\sigma=\rho-\rho_c$ with $\rho_c=0.821$, and extracting $f$, $h$ and $\lambda$ from the fit
(see Fig.\ \ref{figrelax}b).
The resulting parameters were $f=0.712\pm0.003$, $h=0.95\pm0.15$, $\lambda=0.787\pm0.010$. The parameter $\lambda$ is
therefore in agreement with the exponent $\gamma$ extracted from the power law divergence of the relaxation times $\tau_\alpha$.

We have then checked the predictions of MCT on the shape of $\chi_4(t)$, and the divergence of $\chi_4(t^\ast)$ as a function of $t^\ast$
\cite{birbou,twbbb}.
In the Inset of Fig.\ \ref{figchi}, we plot $\chi_4(t)$ on a logarithmic scale.
As shown by the straight lines, during the early and late $\beta$ regime (before it reaches the maximum),
$\chi_4(t)$ can be well fitted by power laws $t^\mu$, with $\mu=0.29$ and $\mu=0.50$, corresponding to the values of $a$ and $b$,
as predicted by MCT.

Finally, in Fig.\ \ref{figtau}b we show the dependence of the maximum $\chi_4(t^\ast)$ as a function of the time $t^\ast$.
The values can be fitted with a power law $\chi_4(t^\ast)\sim {t^\ast}^x$,
with an exponent $x=0.36$, compatible within errors
with the value $1/\gamma$, where $\gamma$ is the exponent of the divergence of times $\tau_\alpha$ and $t^\ast$.
Also in this case, the critical behavior is observed only for $|\rho-\rho_c|<0.005$.

In conclusion, the dynamical behaviour of the model on the random graph is consistent, near the transition, with the Mode Coupling Theory of the
glass transition.
In  particular we have verified the quantitative
prediction which relates the dynamical susceptibility with the decay
of the density-density correlation function.
Note that, due to the simplicity of the model, we have been able to consider a large system size. This is important
because small systems suffer from large finite size effects near the transition, where the asymptotic behavior is expected.

{\em Acknowledgments.} We thank M. Tarzia for very useful discussion.


\begin{thebibliography}{9}
\bibitem{adamgibbs} J.H. Gibbs and E.A. Di Marzio, J. Chem. Phys. {\bf 28}, 373 (1958);
G. Adam and J.H. Gibbs, J. Chem. Phys. {\bf 43}, 139 (1965).
\bibitem{kirkwol} T.R. Kirkpatrick, D. Thirumalai, P. G. Wolynes,  Phys. Rev. A {\bf 40}, 1045 (1989).
\bibitem{cicerone} M.T. Cicerone, F.R. Blackburn, M.D. Ediger, Macromol. {\bf 28}, 8224 (1995); J. Chem. Phys. 102, 471 (1995).
\bibitem{donati} C. Donati, J.F. Douglas, W. Kob, S.J. Plimpton, P.H. Poole and S.C. Glotzer, Phys. Rev. Lett. {\bf 80}, 2338 (1998);
C. Bennemann, C. Donati, J. Baschnagel, S. C. Glotzer, Nature {\bf 399}, 246 (1999).
\bibitem{frapa} Franz and G. Parisi, J. Phys. Cond. Mat. {\bf 12}, 6335 (2000).
\bibitem{fdpg} S. Franz, C. Donati, G. Parisi and S.C. Glotzer, Phil. Mag. B {\bf 79}, 1827 (1999);
C. Donati, S. Farnz, S.C. Glotzer and G. Parisi, J. Non-Cryst. Solids {\bf 307},  215 (2002).
\bibitem{mct} W. G{\"o}tze and L. Sj{\"o}gren, Rep. Prog. Phys. {\bf 55}, 241 (1992).
\bibitem{birbou} G. Biroli and J.-P. Bouchaud, Europhys. Lett. {\bf 67}, 21 (2004).
\bibitem{twbbb} C. Toninelli, M. Wyart, L. Berthier, G. Biroli and J.-P. Bouchaud, Phys. Rev. E {\bf 71}, 041505 (2005).
\bibitem{pspin} T. Kirkpatrick and D. Thirumalai, Phys. Rev. Lett. {\bf 58}, 2091 (1987).
\bibitem{birnew} T. Sarlat, A. Billoire, G. Biroli and J.-P. Bouchaud, {\tt arXiv:0905.3333}.
\bibitem{ulm} M. {Pica Ciamarra}, M. Tarzia, A. de Candia and A. Coniglio, Phys. Rev. E {\bf 67}, 057105 (2003); Phys. Rev. E {\bf 68}, 066111 (2003).
\bibitem{nota1} By ``site'' we mean here a cell having $z$ internal possible positions. In Fig.\ \ref{figlatt}b each plaquette
formed by four positions represent a single site of the random graph.
\bibitem{nfold} A.B. Bortz, M.H. Kalos and J.L. Lebowitz, J. Comp. Phys. {\bf 17}, 10 (1975).
\bibitem{nota_rho} The density where relaxation functions and dynamical susceptibility diverge numerically is found to be different from the dynamical
transition calculated within a 1-step RSB {\em ansatz} analytical solution ($\rho_d=0.808$ \cite{ulm}).
One possible reason for this discrepancy is that more than one step RSB is needed to describe correctly the transition.
\end{thebibliography}
\end{document}